\documentclass[aps,prb,twocolumn]{revtex4-2}

\usepackage{amsmath}
\usepackage{amssymb}
\usepackage{graphicx}
\usepackage{epstopdf}
\usepackage{hyperref}
\usepackage{braket}

\usepackage[version=4]{mhchem}

\begin{document}

\author{Hongsoek Kim}
\affiliation{Department of Physics, Korea Advanced Institute of Science and Technology, Daejeon, Republic of Korea}
\author{Se Kwon Kim}
\affiliation{Department of Physics, Korea Advanced Institute of Science and Technology, Daejeon, Republic of Korea}

\title{Topological phase transition in magnon bands in a honeycomb ferromagnet driven by
sublattice symmetry breaking}

\begin{abstract}
    Ferromagnetic honeycomb systems are known to exhibit a magnonic topological phase under the existence of the next-nearest neighbor Dzyaloshinskii-Moriya interaction (DMI).
    Motivated by the recent progress in the sublattice-specific control of magnetic anisotropy,
    we study the topological phase of magnon bands of honeycomb ferromagnetic monolayer and bilayer with the sublattice symmetry breaking due to the different anisotropy energy in the presence of the DMI.
    We show that there is a topological phase transition between the topological magnon insulator and the topologically trivial magnon phase driven by the change of the relative size of the DMI and the anisotropy differences between the sublattices.
    The magnon thermal Hall conductivity is proposed as an experimental probe of the magnon topology.    
\end{abstract}

\maketitle

\section{Introduction}

A topological order of many-body systems has been a subject of intense investigation in modern condensed matter physics since the discovery of the quantum Hall state~\cite{KlitzingPRL1980}.
The quantum Hall state exhibits the Hall conductivity quantized in integer ratio when the external field is applied to the two-dimensional electron gas, the origin of which is identified as a topological invaraint by the Thouless-Kohmoto-Nightingale-den Nijs (TKNN) formula~\cite{Thouless_1982}.
Later, a novel topological phase called a topological insulator has been suggested and discovered in monolayer graphene and \ce{HgTe} quantum well~\cite{KanePRL2005,BernevigPRL2006}, which is different from the quantum Hall state in that it can exist in the presence of the timer reversal symmetry. These studies of topological properties of two-dimensional electron systems have been expanded to the studies about multilayer materials.
It has been found that manipulation of topological property of electronic system is possible by the stacking~\cite{AvetisyanPRB2010,LinPRB2014,AokiSSC2007}, distance between layers~\cite{OhtaS2006,McCannPRL2006,MinPRB2007,McCannRoPiP2013,JuN2015,MunozPRB2016} or the stacked angle~\cite{CaoN2018,CaoN2018a}.

By extension from the electron system which is fermionic, recently there has been much attention to the topological order in bosonic systems.
Specifically in ordered magnets, a topological magnon insulator has been studied as a bosonic analog of the topological insulator.
In the honeycomb lattice with Dzyaloshinskii-Moriya interaction (DMI)~\cite{DzyaloshinskyJoPaCoS1958,MoriyaPR1960}, the realization of topological magnon insulator has been studied theoretically~\cite{ZhangPRB2013,MookPRB2014,KimPRL2016,OwerreJoAP2016,RueckriegelPRB2018,LeePRB2018} and probed experimentally~\cite{ChenPRX2018,CaiPRB2021}. The topological magnon insulator states have been also identified in many other systems e.g. with magnets with the Kitaev interaction~\cite{JoshiPRB2018,McClartyPRB2018}, 
magnon-magnon interaction~\cite{LuPRL2021},
photoinduced state~\cite{OwerreJoAP2016},
and magnetostatic interaction~\cite{MatsumotoPRB2011,ShindouPRB2013,WangJoAP2021}.
Experimentally probing a topological magnon insulator with the Hall effect is infeasible because magnons carry no electrical charge.
As an alternative probe accessible through transport measurements, the thermal Hall effect in the magnetic system is theoretically suggested~\cite{OnodaPRB2008,KatsuraPRL2010} and observed in the experiment~\cite{OnoseS2010}.

To utilize the topological effects of many-body systems for practical application it is desirable to be able to tune their topological properties.
In particular, magnetic systems allow us to change the magnetic properties by several means such as electric~\cite{HuangNN2018,BurchN2018}, acoustic~\cite{GoPRL2019}, and optical means~\cite{OtaniNP2017}.
We are interested in the effects of the inversion symmetry breaking  in the honeycomb lattice magnet which can be induced by breaking sublattice symmetry.
Multilayer \ce{CrI3} shows varying interlayer coupling by the different stacking states~\cite{SivadasNL2018},
and interlayer coupling can cause sublattice symmetry breaking and affect the sublattice-specific magnetic property~\cite{ChenPRX2021}.
Also, the method of sublattice symmetry breaking by controlling anisotropy energy in van der Waals heterostructure of \ce{CrI3} and transition metal dichalcogenides (TMDs) has been suggested~\cite{HidalgoSacotoPRB2020}.

Previously it has been shown that both DMI and sublattice symmetry breaking can induce a magnon bandgap at high symmetry points.
However, the topological property of the magnon bands with both effects involved has not been studied yet.
In this work, we study a topological phase transition in a honeycomb lattice ferromagnet by tuning the parameters related to DMI and anisotropy energy. Our paper is organized as follows. We discuss the magnon band structure of the monolayer system in Sec.~\ref{sec:mono} and the bilayer system in Sec.~\ref{sec:bi}. We investigate the thermal transport property of both the monolayer and the bilayer in Sec.~\ref{sec:thermal}, and conclude with a summary in Sec.~\ref{sec:summary}.

\section{Monolayer} \label{sec:mono}

In this section, we investigate the effect of sublattice symmetry breaking and the DMI on the magnon bands of a ferromagnetic honeycomb lattice monolayer.

\begin{figure}
    \includegraphics[width=1.0\linewidth]{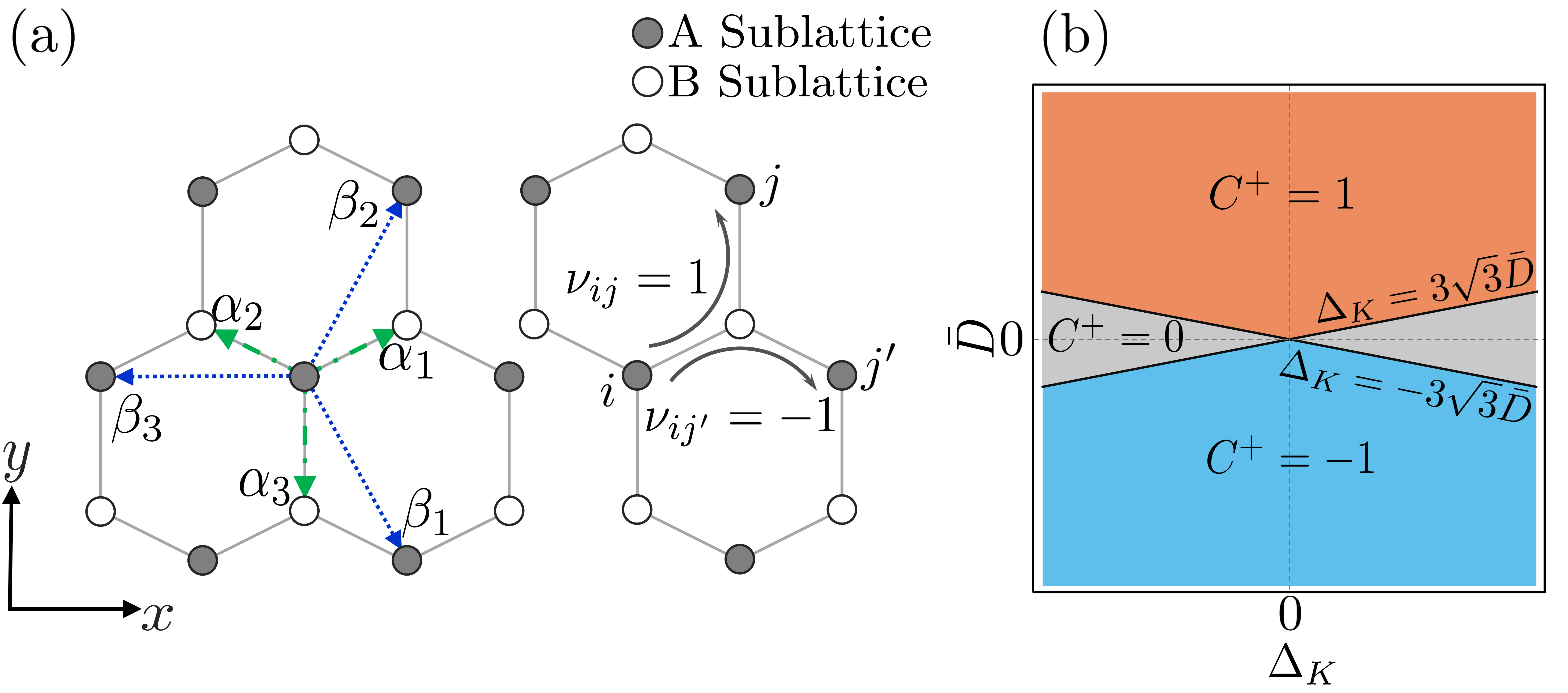}
    \caption{(a) Monolayer honeycomb lattice structure and the direction of the DM interaction $\nu_{ij}$.
    (b) The Chern number of the upper magnon band $C^+$ by the relative size of anisotropy energy difference $\Delta_K$ (between the two sublattices) and the average size of next-nearest-neighbor DMI $\bar{D}$. The topological phase boundaries are given by $\Delta_K = \pm 3 \sqrt{\bar{D}}$.}
    \label{fig:monolayer}
\end{figure}

\subsection{Magnon Bands} \label{sec:magband}

\begin{figure*}[htb]
    \includegraphics[width=1.0\linewidth]{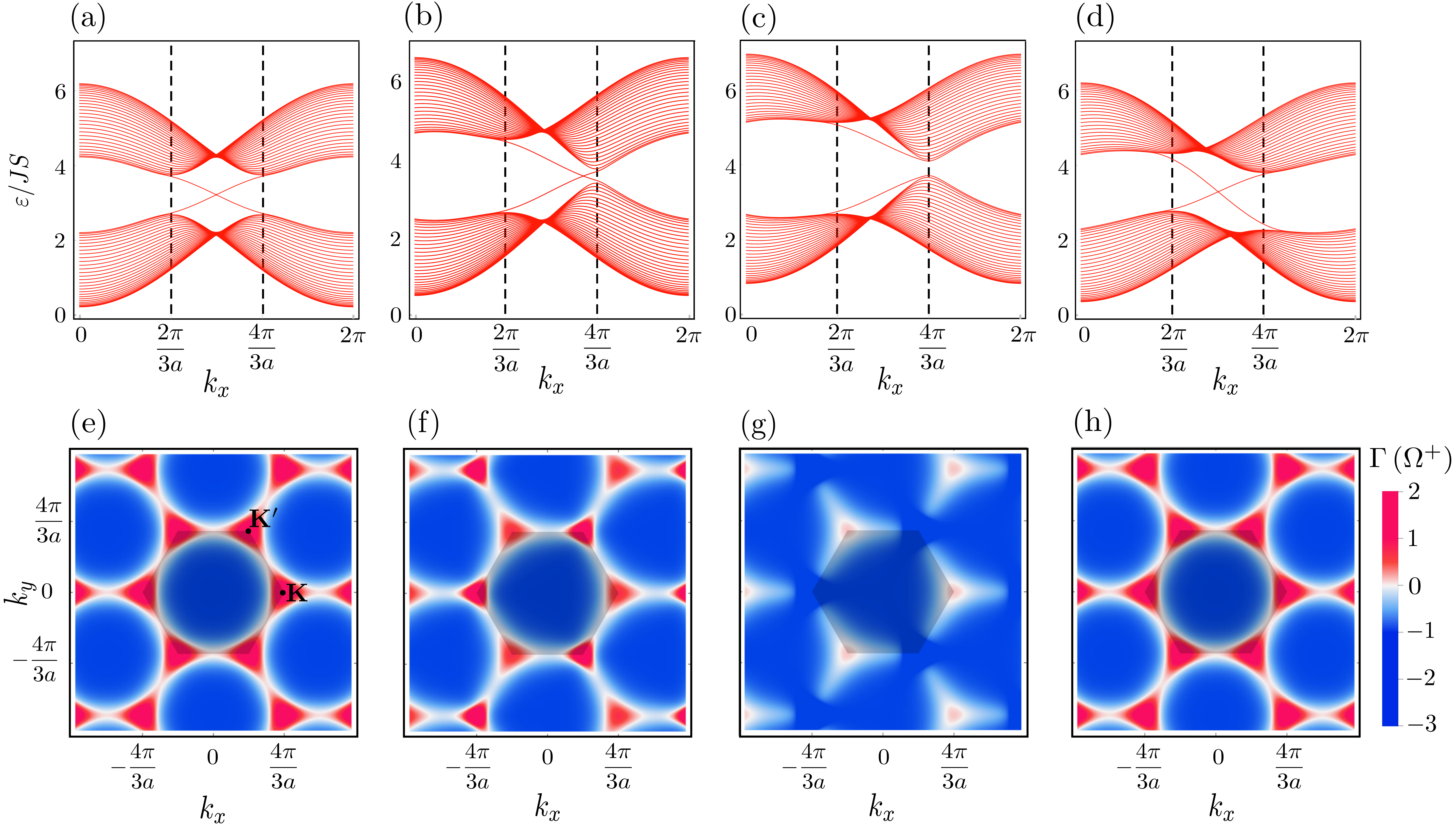}
    \caption{Plots of the band structure (a-d), and the Berry curvature of the upper band in log scale $\Gamma(\Omega^+)=\text{sgn}\left(\Omega^+\right)\log\left(1+\lvert\Omega^+\rvert\right)$  (e-h) for several cases with: 
    (a,e) $\bar{D}=0.1J$ and $\Delta_K=0$ with no difference in the anisotropy energy between sublattices,
    (b,f) $\Delta_K=0.4\bar{D}$, 
    (c,g) $\Delta_K=0.7\bar{D}$,
    when the gap closes at $\Delta_K=3\sqrt{3}\bar{D}\approx 0.52\bar{D}$.
    By comparing (c) and (b), we can see the disappearance of the edge state when $\Delta_K$ increases from $0.4 \bar{D}$ to $0.7 \bar{D}$. By comparing (g) and (f), we can see the inversion of the Berry curvature sign at $K'$ points.
    (d,h) The DMI for sublattices are set to be different as $D_A=0.15J$ and $D_B=0.05J$ when $\Delta_K=0$.
    Note that the DMI difference between sublattices tilts band, but does not change the topology of the magnon band.
    Band structures for each plot are calculated with a ribbon geometry with zigzag termination and 30 unit cells width (60 sites)~\cite{NetoRoMP2009}.}
    \label{fig:monoband}
\end{figure*}

Our system is the ferromagnetic honeycomb lattice monolayer with the DMI and easy-axis anisotropy. 
We first write a magnetic Hamiltonian of the system as~\cite{KimPRL2016,OwerreJoAP2016,HidalgoSacotoPRB2020}:
\begin{equation}
    \begin{split}
    H_{\text{mag}}=&-J\sum_{\langle i,j \rangle}\mathbf{S}_i\cdot\mathbf{S}_j \\
    &+D_A\sum_{\langle\langle i,j \rangle\rangle \in A} \nu_{ij}\hat{\mathbf{z}}\cdot (\mathbf{S}_i\times\mathbf{S}_j)\\
    &+D_B\sum_{\langle\langle i,j \rangle\rangle \in B} \nu_{ij}\hat{\mathbf{z}}\cdot (\mathbf{S}_i\times\mathbf{S}_j)\\
    &-K_A\sum_{i\in A}(S_i^z)^2-K_B\sum_{i \in B}(S_i^z)^2-H\sum_i S_i^z.
    \end{split}
\end{equation}
The first term represents an isotropic Heisenberg interaction between nearest-neighbor (NN) spins with $J>0$.
The nearest-neighbor DMI can exist if the out-of-plane electric field or in-plane magnetic field is applied, but it is not present in our system because the midpoint between neighboring A and B site is the inversion center~\cite{MoriyaPR1960}.
The next-nearest-neighbor (NNN) DMI with coefficients $D_A$ and $D_B$ can exist for A and B sublattices, respectively, and the DMI vector is in the out-of-plane direction which we denote by $\hat{\mathbf{z}}$.
The NNN DMI between different sublattices can be distinct as the sublattice symmetry of our system is broken~\cite{BhowmickPRB2020}.
The sign $\nu_{ij}=1(-1)$ indicates clockwise(counter-clockwise) hopping from $i$ to $j$.
Two terms with $K_A$ and $K_B$ are easy-axis anisotropy energy depending on the sublattice~\cite{HidalgoSacotoPRB2020}. 
The last term is the coupling to external field $H$ in $\hat{\mathbf{z}}$ direction.
Without loss of generality, we assume $H\geq  0$.
A schematic diagram of our system is shown in Fig.~\ref{fig:monolayer}(a).

We consider a small deviation of the spins from the ground state uniformly aligned in $\hat{\mathbf{z}}$ direction.
To this end, we adopt the linear spin-wave approximation. Application of the Holstein-Primakoff transformation~\cite{HolsteinPR1940}
$S_{i,\gamma}^+=S_{i,\gamma}^x+iS_{i,\gamma}^y=(2S-n_{i,\alpha})^{1/2} b_{i,\gamma},\,
 S_{i,\gamma}^-=(S_{i,\gamma}^+)^\dagger, \,
 S_{i,\gamma}^z=S-n_{i,\gamma}$ with $n_{i,\gamma}=b_{i,\gamma}^\dagger b_{i,\gamma}$ yields the following effective magnon Hamiltonian
\begin{equation}
    \begin{split}
    H_{\text{m}}=&(3JS+H)\sum_{i, \gamma} b_{i,\gamma}^\dagger b_{i,\gamma}\\
    &-JS \sum_{\langle i,j \rangle} (b_{i,A}^\dagger b_{j,B} + \text{h.c.})\\
    &+2S \sum_{i,\gamma} K_\gamma  b_{i,\gamma}^\dagger b_{i,\gamma}\\
    &-S \sum_{\langle\langle i,j \rangle\rangle, \gamma} D_\gamma (i\nu_{ij}b_{i,\gamma}^\dagger b_{j,\gamma}+ \text{h.c.}),
    \end{split}
\end{equation}
up to quadratic order in the magnon operators, where $b_{i,\gamma}^\dagger$ is a magnon creation operator on site $i$ of sublattice $\gamma=A,B$ and $b_{i, \gamma}$ is the corresponding magnon annihilation operator. The first term is the on-site energy from Heisenberg interaction and external field coupling; the second term is from the nearest neighbor Heisenberg interaction energy; the third and fourth terms are on-site anisotropy energy and next-nearest-neighbor hopping by the DMI.

By defining a two-component spinor operator $\Psi_\mathbf{k} = (b_{A,\mathbf{k}},b_{B, \mathbf{k}})^T$, we can write the momentum space representation of the magnon Hamiltonian as
\begin{equation}\label{eq:Hk}
    \begin{split}
    H_{\text{m}}=
    \Psi_\mathbf{k}^\dagger \biggl[&
        \Bigl(3JS+H+(K_A+K_B)S+ \Bigr.\biggr. \\\biggl.\Bigl.
        &(D_A-D_B)S\sum_j \sin(\mathbf{k}\cdot \boldsymbol{\beta_j}) \Bigr) \mathbf{I}
        +\mathbf{h}(\mathbf{k})\cdot\boldsymbol{\tau}
        \biggr] \Psi_\mathbf{k},
    \end{split}
\end{equation}
where $\boldsymbol{\tau}$ is a vector of Pauli matrices that represents the sublattice degrees of freedom and
\begin{equation} 
    \mathbf{h}(\mathbf{k})=\left(
    \begin{gathered}
        -JS \, \sum_j\cos[\mathbf{k}\cdot \boldsymbol{\alpha}_j] \\
         JS \,\sum_j\sin[\mathbf{k}\cdot \boldsymbol{\alpha}_j] \\ 
        \Delta_K S+2\bar{D}S\sum_j \sin (\mathbf{k}\cdot \boldsymbol{\beta}_j)
    \end{gathered}\right),
\end{equation}
where $\Delta_K=K_A-K_B$ is the anisotropy difference between sublattices, and $\bar{D}=(D_A+D_B)/2$ is an average value of the DMI coefficients. 
We denote the distances between nearest and next-nearest neighbor as $d$ and $a=\sqrt{3}d$, and
$\boldsymbol{\alpha}_j$, $\boldsymbol{\beta}_j$ are NN and NNN vectors defined in Fig.~\ref{fig:monolayer}(a).
The topology of the system is related to the $\mathbf{k}$ dependence of the eigenstates, and it is determined by the term $\propto \mathbf{h}(\mathbf{k}) \cdot \boldsymbol{\mathbf{\tau}}$ in the Hamiltonian that intertwines the two components of the spinor.
By diagonalizing the Hamiltonian, we obtain the dispersion of the upper (lower) band $E_m^+$ $(E_m^-)$ as
\begin{equation} \label{eq:band}
    \begin{split}
    E_m^\pm=&3JS+H+(K_A+K_B)S\\ &+(D_A-D_B)\sum_j\sin(\mathbf{k}\cdot\boldsymbol{\beta}_j)\pm\lvert \mathbf{h}(\mathbf{k})\rvert.
    \end{split}
\end{equation}

The energy difference between the upper band and the lower band at two high-symmetry points $\mathbf{K}=(4\pi/3a,0)$ and $\mathbf{K'}=(2\pi/3a,2\pi/\sqrt{3}a)$ are given by $|\Delta_K+3\sqrt{3}\bar{D}|S$ and $|\Delta_K-3\sqrt{3}\bar{D}|S$ respectively. 
Therefore the gap on $\mathbf{K}(\mathbf{K}')$ will close at $\Delta_K=-3\sqrt{3}\bar{D}(3\sqrt{3}\bar{D})$ and becomes finite otherwise.
In Fig.~\ref{fig:monoband} (a)-(d), the band is plotted for four different cases that will be explained shortly by calculating the magnon band in a quasi-one-dimensional geometry where a periodic boundary condition is given in $\hat{\mathbf{x}}$ direction and zigzag termination and 30 unit cells in $\hat{\mathbf{y}}$ direction. The Berry curvature of upper($+$) and the lower($-$) band can be calculated according to the formula $\Omega^{\pm}_m=\mp \hat{\mathbf{n}}\cdot(\partial_{k_x} \hat{\mathbf{n}}\times\partial_{k_y} \hat{\mathbf{n}})$ where $\hat{\mathbf{n}}$ is the unit vector along $\mathbf{h}$~\cite{XiaoRoMP2010}.
The corresponding Chern number of the upper($+$) and lower($-$) band is evaluated as $C^{\pm}=(1/2\pi) \int_{B.Z} d^2 k \Omega^{\pm}=\pm 1$~\cite{QiPRB2008}. The Chern number of the upper band is plotted as a function of $\Delta_K$ and $\bar{D}$ in Fig.~\ref{fig:monolayer}(b).

When there is no DMI and only the sublattice symmetry breaking exists ($\bar{D}= 0,\,\Delta_K\neq 0$), the system is shown to be in a topologically trivial phase analogous to an ordinary electronic insulator with no topology, having the trivial Berry curvature with $C=0$ for both upper and lower bands~\cite{HidalgoSacotoPRB2020}.
On the other hand, if there is DMI only and no sublattice symmetry breaking ($\bar{D}\neq 0,\,\Delta_K=0$), it has been known that the system becomes a topological insulator having a gapped band with edge modes, possessing the nontrivial Berry curvature with $C=1$ and $C=-1$ for the two bands~\cite{KimPRL2016,OwerreJoAP2016} [Fig.~\ref{fig:monoband}(a,e)].
If both $\bar{D}$ and $\Delta_K$ are nonzero, the state of the system changes depending on the relative size of $\bar{D}$ and $\Delta_K$.

For $0<\Delta_K <3\sqrt{3}\bar{D}$, the gap is decreased at $\mathbf{K}'$ compared to the $\Delta_K=0$ case shown in Fig.~\ref{fig:monoband}(a) and increased at $\mathbf{K}$ as shown in Fig.~\ref{fig:monoband}(b), while keeping an edge mode connecting the two bulk bands.
When $\Delta_K$ becomes larger than the critical value given by $3\sqrt{3}\bar{D}$, we can see that after the band touching the edge mode disappeared and the gap became finite again in Fig.~\ref{fig:monoband}(c).
The sign of the Berry curvature at $\mathbf{K}'$ flips before and after the band touching [Fig.~\ref{fig:monoband}(f,g)], and and the Chern number of each band becomes trivial for $\Delta_K$ larger than $3\sqrt{3}\bar{D}$ [Fig.~\ref{fig:monolayer}(b)].
The same sign changes of Berry curvature from positive to negative and the gap reopening occur on $\mathbf{K}$ as $\Delta_K$ decreases below $3-\sqrt{3}\bar{D}$.
Different DMIs per sublattices ($D_A\neq D_B$) give a small tilt to the band structure but does not affect the band topology as shown in Fig.~\ref{fig:monoband}(d,h).
These are our first main results: the magnon dispersion in Eq.~(\ref{eq:band}), and the topological phase diagram [Fig.~\ref{fig:monolayer} (b)] determined by $\bar{D}$ and $\Delta_K$. Our results indicate that the topological phase of the honeycomb ferromagnet can be manipulated by tuning the relative size of the difference of the anisotropy energy between sublattices parametrized by $\Delta_K$ and the average DMI parametrized by $\bar{D}$.

\subsection{Magnonic Dirac Equation}
We can investigate the contribution of DMI and broken sublattice symmetry to the bandgap with an effective mass model as applied to the electronic topological insulator~\cite{KanePRL2005}.
By expanding the momentum space Hamiltonian [Eq.~(\ref{eq:Hk})] in the vicinity of $\mathbf{K}$ and $\mathbf{K}'$ in terms of $\mathbf{q}$ defined by $\mathbf{k}=\mathbf{K}(\mathbf{K}')+\mathbf{q}$ with $|\mathbf{q}|d \ll 1$, we can write the effective mass Hamiltonian for magnons as
\begin{equation}
    \mathcal{H}_{\pm}=\Psi^\dagger \biggl[ 
        \frac{3}{2}JS(q_x\sigma_x-q_y\sigma_y\tau_z) 
        +\Delta_KS\sigma_z+3\sqrt{3}\bar{D}S\sigma_z\tau_z
        \biggr] \Psi,
\end{equation}
where $\sigma_\alpha$ are Pauli matrices acting on sublattices, and $\tau_z=\pm 1$ indicates states at the $\mathbf{K}(\mathbf{K}')$ points.
In the absence of both $\Delta_K$ and $\bar{D}$, the magnon band possesses a gapless Dirac-like spectrum around $\mathbf{K}$ and $\mathbf{K}'$ points~\cite{FranssonPRB2016}.
The coefficient in front of the $\sigma_z$ proportional to $\Delta_K$ can be interpreted as a Semenoff-type mass~\cite{SemenoffPRL1984} that has the same sign on $\mathbf{K}$ and $\mathbf{K}'$. The coefficient proportional to $\bar{D}$ in $\sigma_z\tau_z$ term that changes sign between the $\mathbf{K}$ and $\mathbf{K}'$ points is referred to as the Haldane mass~\cite{HaldanePRL1988}. 
These two mass terms proportional to $\sigma_z$ or $\sigma_z\tau_z$ contribute to opening a gap.
By diagonalizing above we obtain
\begin{equation}
    \varepsilon_\pm=\pm\sqrt{(\Delta_K\pm 3\sqrt{3}\bar{D})^2S^2+(9/2)J^2S^2q^2},
\end{equation}
with which we can check the gap closing condition $\Delta_K=\pm 3\sqrt{3}\bar{D}$ agrees with Eq.~(\ref{eq:band}) from Sec.~\ref{sec:magband}.

\section{Bilayer} \label{sec:bi}
\begin{figure}
    \includegraphics[width=1.0\linewidth]{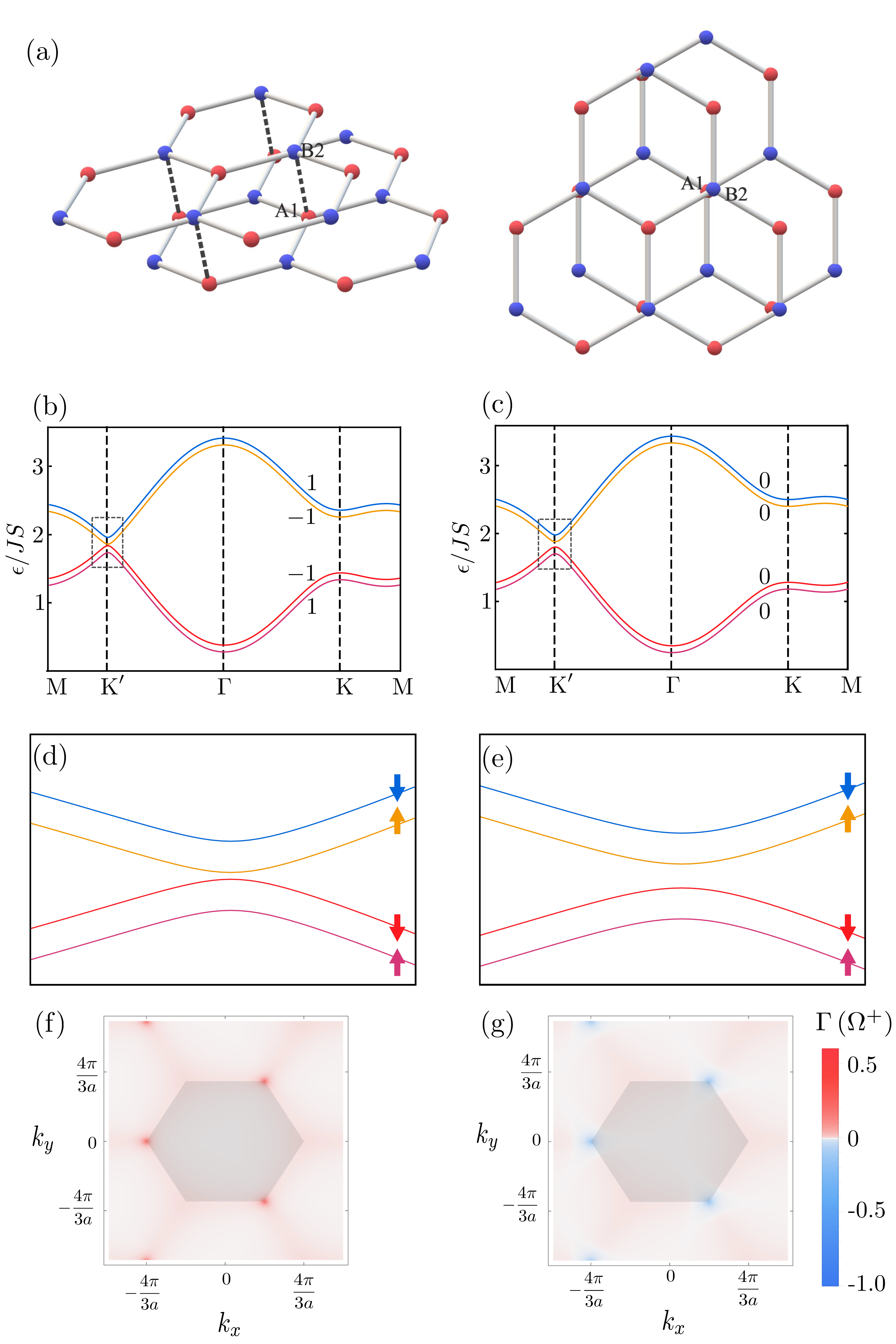}
    \caption{
        (a) Schematic diagram of an AB-stacked bilayer honeycomb lattice. The dark dotted line indicates interlayer coupling.
        The band structure (b,c) and the Berry curvature (f,g) of the bilayer system in two different values of anisotropy, $\Delta_K=0.3J<\Delta_K^c$ (b,d,f) and $\Delta_K=0.6J>\Delta_K^c$ (c,e,g) when the bands touch at $\Delta_K^c \approx 3\sqrt{3}D -0.5J_{12}$.
        The Chern numbers of the four bands are denoted on the corresponding lines.
        Close-up figures of boxed area in (b) and (c) are shown in (d) and (e) with the spin direction of each band shown.
        (f,g) Sign of the Berry curvature on $\mathbf{K}'$ flips at the band touching point between (f) and (g).
        Here we set $\hbar=k_B=J=1$ and $J_{12}=0.2J,\,D=0.1J$ for the calculation.
        }
    \label{fig:bilayer}
\end{figure}

We extend the discussion from the previous section on the monolayer case to the bilayer case. 
Consider an AB-stacked bilayer system consisting of two identical honeycomb ferromagnetic monolayers~\cite{SivadasNL2018}.
A schematic diagram of our system is shown in Fig.~\ref{fig:bilayer}(a). 
Two layers are antiferromagnetically coupled, and therefore spins of each layer will be aligned to the opposite directions ($\hat{\mathbf{z}}$ and $-\hat{\mathbf{z}}$) in the ground state~\cite{SivadasNL2018,OrtmannsPRB2021}.
The Hamiltonian can be written in three parts $H_{\text{m},\text{bilayer}}=H_1+H_2+H_{12}$, where

\begin{align}
    \begin{split}
        H_1=
        &-J\sum_{\langle i,j \rangle}\mathbf{S}_i\cdot\mathbf{S}_j + \sum_{\langle i,j \rangle}\mathbf{D}_{ij}\cdot (\mathbf{S}_i\times \mathbf{S}_j)\\
        &+\sum_{\langle\langle i,j \rangle\rangle \in \alpha, \, \alpha = A, B} \nu_{ij} D_{\alpha} \hat{\mathbf{z}}\cdot (\mathbf{S}_i\times\mathbf{S}_j)\\
        &-K_{1,A}\sum_{i\in A}(S_i^z)^2-K_{1,B}\sum_{i \in B}(S_i^z)^2-H\sum_i S_i^z
    \end{split}	\label{eq:H1}
    \\[2ex]
    \begin{split}
        H_2=
        &-J\sum_{\langle i,j \rangle}\mathbf{S}_i\cdot\mathbf{S}_j + \sum_{\langle i,j \rangle}\mathbf{D}_{ij}\cdot (\mathbf{S}_i\times \mathbf{S}_j)\\
        &+\sum_{\langle\langle i,j \rangle\rangle \in \alpha, \, \alpha = A, B} \nu_{ij} D_{\alpha} \hat{\mathbf{z}}\cdot (\mathbf{S}_i\times\mathbf{S}_j)\\
        &-K_{2,A}\sum_{i\in A}(S_i^z)^2-K_{2,B}\sum_{i \in B}(S_i^z)^2-H\sum_i S_i^z
    \end{split}	\label{eq:H2}
    \\[2ex]
    H_{12}&=J_{12}\sum_{(i,j)} \mathbf{S}_i\cdot \mathbf{S}_j.	\label{eq:H12}
\end{align}
Here, $H_1$ and $H_2$ are the intralayer Hamiltonian for lower (1) and upper (2) layer.
Since two layers are AB-stacked, effects of anisotropy appear oppositely between the layer, i.e. $K_{1,A}=K_{2,B},\, K_{1,B}=K_{2,A}$. 
The second terms of $H_1$ and $H_2$ are the NN DMIs with the in-plane DM vectors $\mathbf{D}_{ij}$, which can be present from the symmetry considerations due to the AB-stacking of the bilayer and the resultant broken inversion symmetry but does not contribute to the magnon band in our calculations performed below.
The last term in the Hamiltonian $H_{12}$ is the interlayer coupling term, $J_{12} > 0$ is the interlayer antiferromagnetic exchange coupling, and $(i,j)$ is the interlayer nearest neighbor. To study the magnon-band topology in the bilayer system below, we assume that the DMI parameters are the same on each sublattice $D \equiv D_A = D_B$ because the difference between the sublattices has no significant effect on the topological properties of magnon bands as shown for the monolayer case above. By applying the Holstein-Primakoff transformation to $H_{\text{m},\text{bilayer}}$ and defining the momentum space operators $b_{\mathbf{k},\gamma}$ where $\gamma=1A,\,1B,\,2A,\,2B$ indicates layer(1,2) and sublattice(A,B), we can write down the momentum space Hamiltonian similar to the previous section:
\begin{equation}
    H(\mathbf{k})=\psi_\mathbf{k}^\dagger\begin{pmatrix} \mathcal{A}(\mathbf{k})&\mathcal{B}\\\mathcal{B}&\mathcal{A}(\mathbf{k}) \end{pmatrix}\psi_\mathbf{k},
\end{equation}
where $\mathcal{A}(\mathbf{k})$ and $\mathcal{B}$ are certain $4\times 4$ matrices. See the Appendix~\ref{app:bilayer} for the definition of these matrices. Here, $\psi_\mathbf{k}=(\phi_\mathbf{k},\phi_{-\mathbf{k}}^\dagger)^T$ with $\phi_\mathbf{k}=(b_{\mathbf{k},1A},b_{\mathbf{k},1B},b_{\mathbf{k},2A},b_{\mathbf{k},2B})$. 

We obtain four bands from the numerical diagonalization, two for each layer.
By tuning the value of $\Delta_K$ we can change the gaps between two bands from the same layer.
The magnon bands in the bilayer system reacts to the increase of $\Delta_K$ in a similar way to the monolayer case; gaps decrease to zero as $\Delta_K$ increase to a critical value and increase again after the critical point.
The critical value of $\Delta_K$ at which the band gap closes is shifted from the monolayer case $3\sqrt{3}D$ by the interlayer coupling $J_{12}$ as it effectively serves as an easy-axis anisotropy albeit from a different origin.

The Berry curvature is calculated numerically using the formula
$\Omega_{n}(\mathbf{k}) = 
-2 \text{Im} \sum_{m\neq n} 
(\bra{u_{n\mathbf{k}}} \partial_{k_x}H \ket{u_{m\mathbf{k}}} \bra{u_{m\mathbf{k}}} \partial_{k_y}H \ket{u_{n\mathbf{k}}})/
[\epsilon_m(\mathbf{k})-\epsilon_n(\mathbf{k})]^2$
where $\ket{u_{n\mathbf{k}}}$ is the eigenvector obtained from the diagonalization.
Before and after the band closing, the Berry curvature of each band shows a sign change at the boundary of the Brillouin zone [Fig.~\ref{fig:bilayer}(f,g)], and Chern numbers of each band change from $+1$ and $-1$ to zeros [Fig.~\ref{fig:bilayer}(b,c)]. 
This confirms that the topological phase transition through tuning parameters $\Delta_K$ and $D$ occurs also in the bilayer system.

\section{Thermal Hall effect} \label{sec:thermal}

\begin{figure}
    \includegraphics[width=1.0\linewidth]{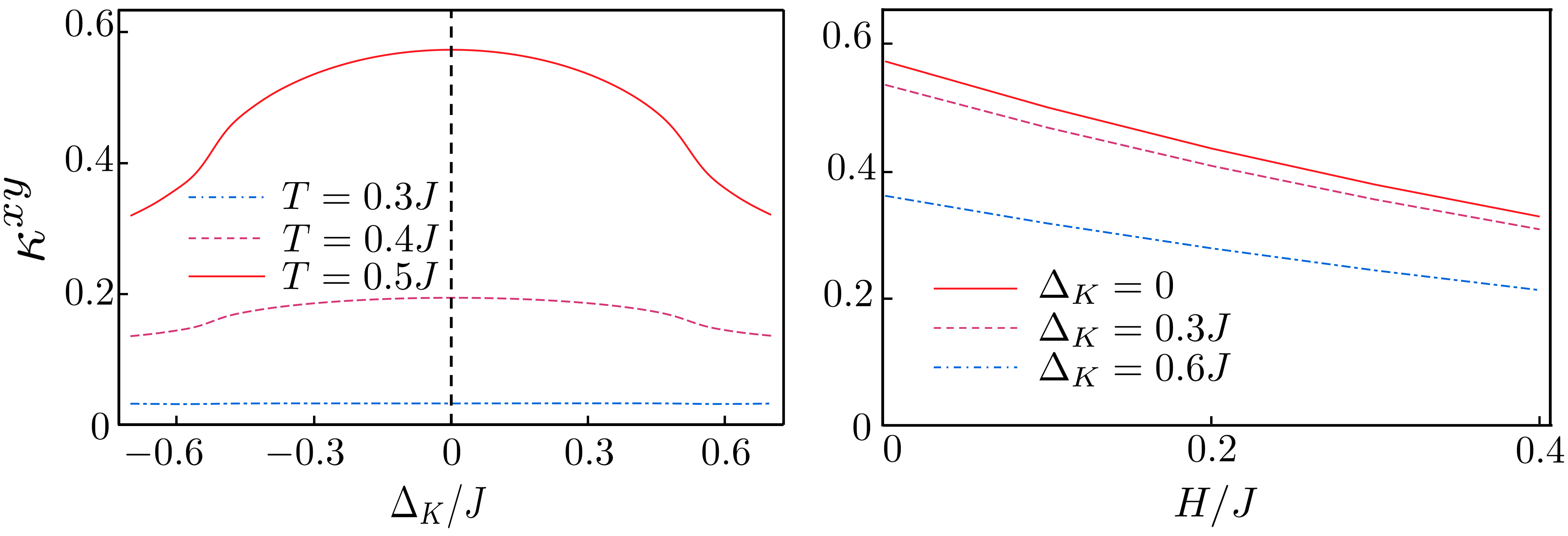}
    \caption{Thermal Hall conductivity $\kappa^{xy}$ of the monolayer honeycomb lattice ferromagnet. (a) The anisotropy energy difference between sublattices decreases $\kappa^{xy}$. Higher temperature supports larger $\kappa^{xy}$ by increasing the number of thermal magnons.
    No external field is applied for this calculation.
    (b) External field decreases $\kappa^{xy}$ by increasing the magnon energy and thereby suppressing the number of thermal magnons.
    We set $\hbar=k_B=J=1$ and used value of $D_A=D_B=0.1J$ for both calculation.}
    \label{fig:monoTH}
\end{figure}

\begin{figure}
    \includegraphics[width=1.0\linewidth]{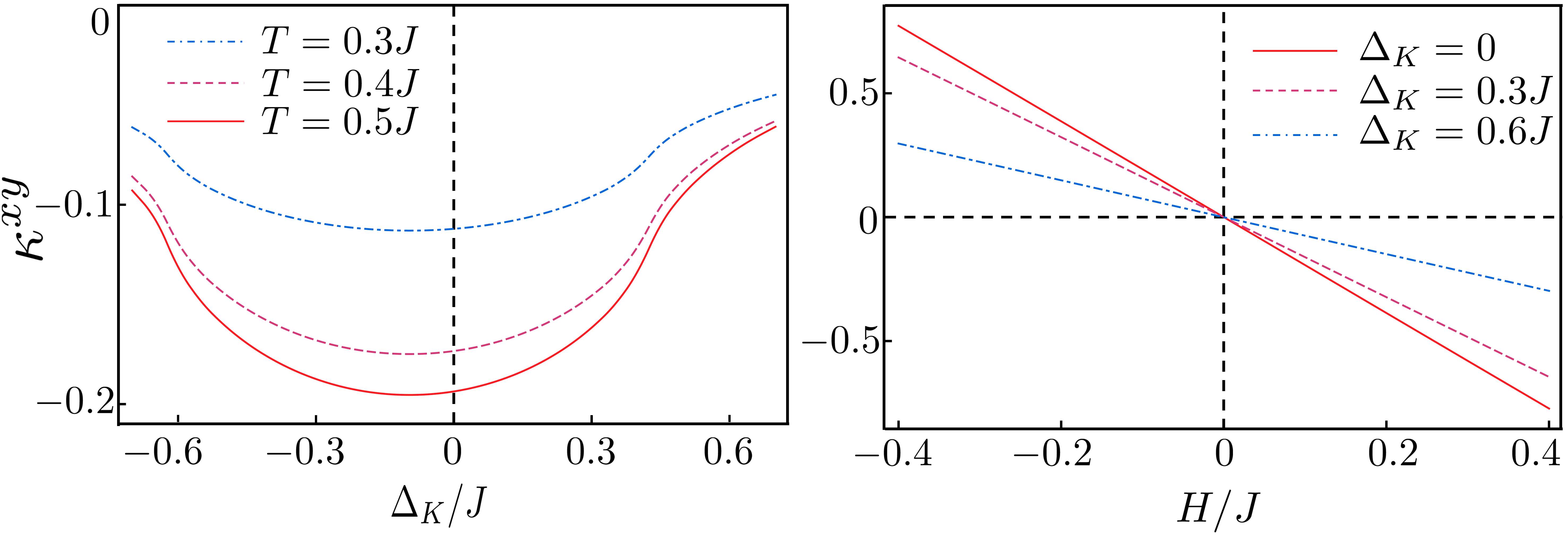}
    \caption{Thermal Hall conductivity $\kappa^{xy}$ of the bilayer honeycomb magnet.
    (a) Anisotropy difference $\Delta_K$ decreases size of the conductivity. 
    The thermal Hall conductivity is larger in higher temperatures due to the larger number of thermal magnons.
    Note that the thermal Hall conductivity is not symmetric under the sign change of $\Delta_K$, which is due to the interlayer coupling that serve as an effective sublattice-dependent easy-axis anisotropy for magnons.
    (b) Two monolayers have opposite spin ground state and react to the external field oppositely. 
    The thermal Hall conductivity vanishes without out-of-plane magnetic field that breaks time-reversal symmetry and increases as the strength of the magnetic field increases.
    In the calculation we set $\hbar=k_B=J=1$ and used value of $D_A=0.1J$, $J_{12}=0.2J$ for both calculation, and $H=0.1J$ for (a).}
    \label{fig:biTH}
\end{figure}

The effects of the DMI and the broken sublattice symmetry on a magnon band can be probed by the thermal Hall conductivity $\kappa^{xy}=-\partial J_Q^x/\partial_y T$ parametrizing a transverse heat current induced by a temperature gradient, both in monolayer and bilayer cases. To calculate the thermal Hall conductivity we use a formula $\kappa^{xy}=-(2k_B^2T/\hbar V)\sum_{n, \mathbf{k}\in B.Z.} \Omega^n(\mathbf{k})c_2(\rho_n)$ derived in Ref.~\cite{MatsumotoPRB2011}, 
where $c_2(\rho)=(1+\rho)(\log \frac{1+\rho}{\rho})^2-(\log\rho)^2-2\text{Li}_2(-\rho)$, $\text{Li}_2$ is the polylogarithmic function, $\rho_n=[\exp(\beta E_n)-1]^{-1}$ is the Bose-Einstein distribution function with $\beta = k_B T$, and $V$ is the sample area.
The results of the calculation for the monolayer and the bilayer are shown in Fig.~\ref{fig:monoTH} and Fig.~\ref{fig:biTH}, respectively.
In the numerical calculation, we set $\hbar=k_B=J=1$ and used value of $D_A=D_B=0.1J$. 
Constant value $H=0.3J$ of the external field is used in calculating the effects of temperature and $\Delta_K$ ((a) of Fig.~\ref{fig:monoTH} and Fig.~\ref{fig:biTH}).

As it can be seen from the calculated Chern numbers, anisotropy energy difference $\Delta_K$ competes with the DMI effect,
and large $\Delta_K$ reduces the accumulative sum of the Berry curvature over the Brillouin zone, of each band.
Therefore, when $\bar{D}\neq 0$ and $\Delta_K = 0$, the system exhibits the nonzero thermal Hall conductivity and its magnitudes decreases as the size of $\Delta_K$ becomes larger [Fig.~\ref{fig:monoTH}(a)].
Interlayer coupling causes a bias to $\Delta_K<0$ in bilayer case as it did to the band touching point [Fig.~\ref{fig:biTH}(a)].

Temperature increases the number of thermally populated magnons through the Bose-Einstein distribution, and the larger number of carrier exhibits larger thermal Hall conductivity.
In monolayer, the existence of the external magnetic field $H$ decreases the conductivity, as $H$ parallel to the direction of spin alignment increases the overall magnon energy [Fig.\ref{fig:monoTH}(b)].
In the bilayer system, nonzero external field $H$ is essential in observation of the thermal Hall conductivity as it explicitly breaks the time-reversal symmetry.
In the absence of an external field $H=0$, the thermal Hall effects from two layers with opposite spin alignment cancel out each other perfectly.
Similar to the monolayer case, the external field decreases magnon energy of spins in a parallel direction and increases magnon energy of spins in antiparallel direction.
Therefore when the magnon energy of spins antiparallel to the external field dominates, and $\kappa^{xy}$ indicates an opposite sign with $H$ and its size increases as $H$ becomes larger.
This result is shown in Fig.~\ref{fig:biTH}(b).
In addition to the aforementioned thermal Hall effect, the spin Nernst effect~\cite{WangJoPDAP2018,KovalevPRB2016} is also expected to be present in our system.
For the spin Nernst effect, the external magnetic field is not necessary in the bilayer system because spin wave generated from each layer add up even when the system has a time-reversal symmetry.\\

\section{Summary} \label{sec:summary}

To sum up, we have compared two factors that contribute to the gap opening of the Dirac-like honeycomb ferromagnet: the next-nearest neighbor DMI and the sublattice symmetry breaking through the magnetic anisotropy energy.
To this end, we have constructed the spin Hamiltonian of mono- and bilayer honeycomb ferromagnet and adapted the magnon picture to study its collective excitations.
The band structure and the Chern number of magnons are calculated that can specify the topological state.
Then, we have investigated how two factors compete with each other. 
The system becomes a topological magnon insulator when the DMI is dominant, and it exhibits a phase transition to a topologically trivial magnonic insulator as the anisotropy energy difference per sublattice exceeds certain critical value $|\Delta_K|=3\sqrt{3}\bar{D}$.
This value is slightly shifted in the bilayer case by the interlayer coupling acting on different sublattice depending on layers.
As an experimental realization, we suggested the thermal transport related to the magnon Berry curvature.
The size of the thermal Hall current has been shown to be tunable through the relative size of DMI and the anisotropy energy difference.

\begin{acknowledgments}
This work was supported by Brain Pool Plus Program through the National Research Foundation of Korea funded by the Ministry of Science and ICT (NRF-2020H1D3A2A03099291), by the National Research Foundation of Korea(NRF) grant funded by the Korea government(MSIT) (NRF-2021R1C1C1006273), and by the National Research Foundation of Korea funded by the Korea Government via the SRC Center for Quantum Coherence in Condensed Matter (NRF-2016R1A5A1008184).
\end{acknowledgments}

\begin{appendix}

\section{Magnon Hamiltonian in the honeycomb bilayer}
\label{app:bilayer}

By applying the Holstein-Primakoff transformation to $H_\text{m,bilayer} = H_1 + H_2 + H_{12}$, where $H_1$, $H_2$, and $H_{12}$ are defined in Eq.~(\ref{eq:H1}), Eq.~(\ref{eq:H2}), and Eq.~(\ref{eq:H12}), respectively, and defining the momentum space operators $b_{\mathbf{k},\gamma}$ where $\gamma=1A,\,1B,\,2A,\,2B$ indicates layer(1,2) and sublattice(A,B), we can write down the momentum space Hamiltonian for the magnons in the bilayer honeycomb system:
\begin{equation}
    H(\mathbf{k})=\psi_\mathbf{k}^\dagger\begin{pmatrix} \mathcal{A}(\mathbf{k})&\mathcal{B}\\\mathcal{B}&\mathcal{A}(\mathbf{k}) \end{pmatrix}\psi_\mathbf{k},
\end{equation}
here $\mathcal{A}(\mathbf{k})$ and $\mathcal{B}$ are $4\times 4$ matrices
\begin{widetext}
\begin{gather}
    \mathcal{A}(\mathbf{k})=
         \begin{pmatrix}
             \begin{bmatrix}
                 v_J+v_H+v_K+v_{12} & 0\\
                 0 & v_J+v_H+v_K
             \end{bmatrix}
             + \frac{1}{2} \mathbf{h}\cdot \boldsymbol{\tau}  & 0\\
             0 & 
             \begin{bmatrix}
                v_J-v_H+v_K & 0\\
                 0 & v_J-v_H+v_K+v_{12}
             \end{bmatrix}
             + \frac{1}{2} \mathbf{h}' \cdot \boldsymbol{\tau}
         \end{pmatrix},
    \\[2ex]
    \mathcal{B}=
    \begin{pmatrix}
        0 & 0 & 0 & v_{12}\\
        0 & 0 & 0 & 0\\
        0 & 0 & 0 & 0\\
        v_{12} & 0 & 0 & 0
    \end{pmatrix},
\end{gather}
\end{widetext}
where $v_J=3JS/2,\, v_K=(K_A+K_B)S/2,\, v_H=HS/2,\, v_{12}=J_{12}S/2$,

\begin{eqnarray}
    \mathbf{h}&=\left(
        \begin{gathered} 
            -JS \, \sum_j\cos[\mathbf{k}\cdot \boldsymbol{\alpha}_j] \\
             JS \,\sum_j\sin[\mathbf{k}\cdot \boldsymbol{\alpha}_j] \\ 
            \Delta_K S+2DS\sum_j \sin (\mathbf{k}\cdot \boldsymbol{\beta}_j)
        \end{gathered}\right),\\
    \mathbf{h}'&=\left(
        \begin{gathered} 
            -JS \, \sum_j\cos[\mathbf{k}\cdot \boldsymbol{\alpha}_j] \\
             JS \,\sum_j\sin[\mathbf{k}\cdot \boldsymbol{\alpha}_j] \\ 
            -\Delta_K S-2DS\sum_j \sin (\mathbf{k}\cdot \boldsymbol{\beta}_j)
        \end{gathered}\right),
\end{eqnarray}
and $\psi_\mathbf{k}=(\phi_\mathbf{k},\phi_{-\mathbf{k}}^\dagger)^T$ for $\phi_\mathbf{k}=(b_{\mathbf{k},1A},b_{\mathbf{k},1B},b_{\mathbf{k},2A},b_{\mathbf{k},2B})$.

This Hamiltonian can be diagonalized with Colpa's method~\cite{ColpaPASMaiA1978}, using a para-unitary transformation to bosonic operators $\chi_\mathbf{k}=\mathfrak{T}\psi_\mathbf{k}$ .
The transformation matrix $\mathfrak{T}$ satisfies $\mathfrak{T}^\dagger \mathfrak{J} \mathfrak{T}=\mathfrak{J}$ , where $\mathfrak{J}=\text{diag}(\mathbf{I}_{4\times 4},-\mathbf{I}_{4\times4})$.
Denoting the matrix part as $\mathfrak{D}$, the Hamiltonian is diagonalized as
\begin{equation}
    H
    = \psi_\mathbf{k}^\dagger \mathfrak{D} \psi_\mathbf{k}
    = \psi_\mathbf{k}^\dagger \mathfrak{T}^\dagger (\mathfrak{T}^\dagger)^{-1} \mathfrak{D} \mathfrak{T}^{-1} \mathfrak{T}\psi_\mathbf{k}
    = \chi_\mathbf{k}^\dagger \mathcal{E} \chi_\mathbf{k},
\end{equation}
and the diagonal matrix $\mathcal{E}$ has a form of $\mathcal{E}=(\mathfrak{T}^\dagger)^{-1} \mathfrak{D} \mathfrak{T}^{-1}=\hbar\,\text{diag}(\omega_1,\dots,\omega_4,\omega_1,\dots,\omega_4)$.

\end{appendix}

\bibliography{proj1manuscript} 

\end{document}